# Management of the Orbital Angular Momentum of Vortex Beams in a Quadratic Nonlinear Interaction

F.A.Bovino*
Quantum Optics Lab - Selex SI
Via Puccini 2 – 16154 Genova - ITALY

M.Braccini, M.Bertolotti, C.Sibilia
Dipartimento di Scienze di Base e Applicate all'Ingegneria
Università di Roma "La Sapienza"
Via Scarpa 16 – 00161 –Roma –Italy
Tel: +390649916942

**\*Corresponding author:** fbovino@selex-si.com

## Abstract

*Light intensity control of the orbital angular momentum of the fundamental beam in a quadratic nonlinear process is theoretically and numerically presented. In particular we analyzed a seeded second harmonic generation process in presence of orbital angular momentum of the interacting beams due both to on axis and off axis optical vortices. Examples are proposed and discussed.*

**Keywords:** Nonlinear optics, Second harmonic generation, Optical Vortex, Orbital Angular Momentum.

## 1. Introduction

In several cases real waves experience jumps in their phase so that they are characterized by the presence of a phase defect. This can lead to an indetermination (singularity) of the phase itself due to a tear of the wave front. The center of the phase defect belongs to a continuous line in space of zero amplitude and the phase circulates around the line creating a vortex. These singularities were studied first by Nye and Berry [1] and the crystallographic term "wave-front dislocation" was introduced to identify them: a phase dislocation is defined as the locus of zero amplitude of a field. Like in crystallography, dislocations can be classified in edge, screw and mixed edge-screw type [1]. An interesting class of optical vortices come from screw wave-front dislocations: the main feature of this kind of singularities is a helical wave-front structure which develops around the dislocation line.

Helical beams are characterized by the presence of angular momentum [2]; under paraxial approximation, it manifests two components: spin, related to the polarization, and orbital, bound to the tangential component of the Poynting vector.

Orbital angular momentum of light plays an important role in many applications, ranging from particle manipulation [3] to quantum information [4], thus it is important to achieve an accurate control of it.

Under paraxial approximation the optical orbital angular momentum (OAM) must be constant during the propagation [5]. With respect to the transverse plane the angular momentum density is defined as $\mathbf{L} = \mathbf{r} \times \mathbf{P}$, being P the linear momentum density of the beam. If we consider a field propagating along $z$ with a transverse distribution $u(\mathbf{r})$, where $\mathbf{r}$ is the generic transverse position coordinate, it can be seen that $\mathbf{L}$ depends only on the azimuthal component of the linear momentum vector [2,6].

Typical beams carrying OAM are Laguerre – Gaussian (LG) beams, as it was shown by Allen et al. [6]. These are modes characterized by the integer numbers $l$ and $p$, which are respectively the azimuthal and the radial indices. In these beams the index $l$ corresponds to the so called topological charge, defined as the winding number of the phase around the singularity; it was shown [6] that the OAM per photon of a LG mode is $l\hbar$. The parameter $p$ denotes the number of nodal rings in the radial direction.

A lot of work has been done to study the propagation/interaction dynamics induced by the superposition of optical vortices [7-9]. As it was widely shown [10-12], nonlinear interaction of vortex beams is a further interesting tool which should be considered in order to control the dynamics of optical vortices (OVs). A variety of interesting effects has been analyzed and observed such as the rising of vortex solitons [13], splitting of solitons [14], and creation and annihilation of vortices due to the interaction of the singularities in nonlinear media by means of the dislocations' dynamics [10,12].

Some of the nonlinear processes which affects the propagation dynamics of optical vortices are parametric frequency conversions in quadratic nonlinear media [15]. Among them, second harmonic generation in a collinear scheme was also copiously studied, using a LG mode as fundamental field (FF) and focusing the attention on the conservation of the orbital angular momentum [17-19]. As we mentioned before, one of the most intriguing applications of vortex beams' interactions is the possibility of getting control not only on the vortices and their position, but also on the field distribution, therefore on the OAM.

The aim of this work is to study the behaviour of the OAM in a seeded second harmonic process, in order to tune the output angular momentum at a suitable value; this task can be accomplished by making use of beams with different values of the OAM and controlling their intensity levels. While integer values of OAM can be obtained by simply nesting an optical vortex on the axis of a Gaussian beam, fractional OAMs are obtained by enclosing in the beam a mixed screw-edge dislocation or by displacing from the beam's axis an OV with integer topological charge. It is interesting to investigate the evolution of fields carrying both integer and fractional values of OAM, as analytically described in ref. [20], because of their manifold applications, which can be found in particular in quantum information, aiming at the realization of n-dimensional quantum states [21]. For this purpose we present a theoretical and numerical analysis of the nonlinear interaction of beams carrying both integer and fractional values of OAM by parallel solving the coupled equations and by means of a Beam Propagation Method (BPM). The interaction occurs in a quadratic nonlinear medium, where a seeded SHG process is considered in both conditions of phase matched and non phase matched interaction. We show that is possible to control the OAM by varying the intensity level of the second harmonic beam.

In what follows the theory of the orbital angular momentum is introduced, in order to predict its behaviour in the quadratic process; then a general treatment of the nonlinear process of seeded SHG through the analytical solution of the nonlinear equations is given; finally, in section 4, we present the numeric method used to solve the equations. The results of some numerical simulations are discussed.

## 2. Orbital angular momentum

As pointed out before, the angular momentum density of a electromagnetic field has its general definition in $\mathbf{r} \times \mathbf{P}$, which implies, in paraxial approximation for a linearly polarized field, that only the orbital component is present and that the nonzero component of the latter is in the $z$ direction. Considering a field with a transverse profile $u(\rho,\varphi)$ in cylindrical coordinates, the OAM per photon, given in $\hbar$ units and calculated with respect to the $z$ – axis (i.e. the propagation axis) becomes:

$$\ell = -i \frac{\int u^*(\rho,\varphi) \frac{\partial u(\rho,\varphi)}{\partial \varphi} \rho d\rho d\varphi}{\int |u(\rho,\varphi)|^2 \rho d\rho d\varphi} = \frac{L}{W}, \qquad (1)$$

where $L$ is the total angular momentum of the field and $W$ is the field's energy.

This general result can also be used to calculate the OAM of a linear superposition of two or more fields and in particular, for our purposes, it is helpful in the case of a field given by the sum of two beams with different angular momenta $\ell_1 = l$ and $\ell_2 = m$. We thus consider a field profile

$$u(\rho,\varphi) \propto A_1 e^{il\varphi} - iA_2 e^{im\varphi}. \qquad (2)$$

Substituting Eq. (2) in Eq. (1), after some simple algebra, an expression of the total orbital angular momentum per photon $\ell$ can be found:

$$\ell = \frac{L_1 + L_2 + A_1 A_2 \left(m e^{i\pi(m-l)} - l e^{i\pi(m-l)}\right) \frac{\sin[\pi(m-l)]}{(m-l)}}{W_1 + W_2 + 2|A_1||A_2| \frac{\sin^2[\pi(m-l)]}{(m-l)}}. \qquad (3)$$

The latter, for integer values of $l$ and $m$ or for $l = m$, depends only on the total angular momenta of the two fields $L_1$, $L_2$ and on the fields' energies $W_1$ and $W_2$:

$$\ell = \frac{L_1 + L_2}{W_1 + W_2}, \qquad (4)$$

It is easy to see that, in these conditions, the overall OAM is $(\ell_1+\ell_2)/2$ when the fields have the same amplitude; on the contrary it reduces to the OAM of the field with the dominant amplitude.

The result of equation (4) will be applied in the next section to the solution of the equations which describe the nonlinear interaction in order to predict the OAM's behaviour.

As mentioned before, in our analysis we consider fields with integer and fractional values of angular momentum: they can be easily modelled by means of Laguerre - Gaussian beams. These modes are a set of helical beams: is thus possible to give an expression of the fractional OAM beam using a modal decomposition in LG beams [26]:

$$u(\rho,\varphi) = \sum_{l=-\infty}^{\infty} \sum_{p=0}^{\infty} C_{lp} \psi_{lp}(\rho,\varphi). \tag{5}$$

Where $\psi_{lp}$ are the LG modes and $C_{lp}$ are the corresponding coefficients which account for the displaced OV.

Substituting eq. (5) in Eq. (1), due to the orthonormal nature of LG modes, it is possible to give an analytical expression of the OAM per photon of the field with fractional momentum:

$$\ell = \sum_{l=-\infty}^{\infty} \sum_{p=0}^{\infty} l \left| C_{lp} \right|^2. \tag{6}$$

## 3. Seeded Second Harmonic Generation

Nonlinear interactions with second-order nonlinearities have been considered as an alternative to intensity dependent changes induced in cubic nonlinear materials [21] for at least two reasons: speed, and, primarily, because of low-loss properties, in contrast with third - order processes. The possibility of lossless operation is attractive, because it can help to revive the idea of fast control of light-by-light for all-optical signal processing. A number of devices has been proposed, like a way of controlling the reflectivity of a fundamental field (FF) beam [23] under conditions of a SH beam more intense than FF, and under suitable phase matching conditions.

Vortex dynamics in presence of phase screw dislocations in a beam and nonlinear interaction, have been also studied [12-14]. In reference [12] a parametric interaction in a quadratic nonlinear medium has been reported under the fundamental field (FF) weak depletion regime; it has been shown that by changing the relative amplitudes and phases of the initial fields, a control of the vortex dynamic can be performed. However nothing is reported about the control of the OAM carried by each beam.

In what follows, we consider beam profiles carrying OAM both for fundamental (FF) and Second Harmonic frequency (SH), both co-propagating in a nonlinear crystal under two different conditions, i.e. phase matching and non phase matching conditions. We perform our analysis combining fields carrying angular momentum with different input amplitudes and we search for the more suitable conditions for controlling the OAM carried by the beams. Considering two fields at the input plane of the nonlinear crystal at frequency ω and 2ω, they must satisfy the nonlinear wave equation; applying the slowly varying envelope approximation, the following equations are found for FF and SH fields:

$$2ik_1 \frac{\partial A_1}{\partial z} + \nabla_T^2 A_1 = -\frac{2\omega_1^2}{c^2} d^{(2)} A_2 A_1^* e^{i\Delta kz}$$
$$2ik_2 \frac{\partial A_2}{\partial z} + \nabla_T^2 A_2 = -\frac{\omega_2^2}{c^2} d^{(2)} |A_1|^2 e^{-i\Delta kz} \quad (7)$$

where $A_1$ and $A_2$ are the complex envelopes of the interacting fields, which depend both from $z$ and the transverse coordinates, $d^{(2)}$ is the nonlinear coefficient, $\omega_1$, $\omega_2$ and $k_1$, $k_2$ are respectively the frequencies and the wave vector of the FF and SH. We solve this problem in phase matching conditions ($\Delta k = 0$) and in conditions of intense SH: we then apply the undepleted approximation for the second harmonic field, thus considering $A_2$ constant with $z$. It is possible to decompose the envelopes in a part with only a $z$ dependence and in a transverse one which accounts for the OAM through a helical phase. Neglecting the radial dependence the envelopes become:

$$A_1(\mathbf{r}, z) = \mathcal{A}_1(z) u(\rho, \varphi) = \mathcal{A}_1(z, \varphi) e^{il\varphi}$$
$$A_2(\mathbf{r}, z) = \mathcal{A}_2 u(\rho, \varphi) = \mathcal{A}_2 e^{im\varphi} \quad (8)$$

In this way we reduced our set of equations to only the first of Eqs. (7) which takes the form of:

$$\frac{\partial \mathcal{A}_1(z, \varphi)}{\partial z} = -i\left[\frac{l^2}{2k_1 r^2} \mathcal{A}_1(z) - \frac{2\omega_1}{n_1 c} d^{(2)} \mathcal{A}_2 \mathcal{A}_1^*(z) e^{i(m-2l)\varphi}\right]. \quad (9)$$

Equation (3) can be solved by separating the real part of $\mathcal{A}_1$ from its imaginary part. After some simple algebraic passages it is found that the solution for the complex envelope of the FF field shows an hyperbolic dependence from the propagation coordinate z.

$$\mathcal{A}_1(z, \varphi) = \mathcal{A}_1(0, \varphi) \cosh\left(\frac{2\omega d^{(2)} |\mathcal{A}_2|}{nc} z\right) + i\mathcal{A}_1^*(0, \varphi) e^{i(m-2l)\varphi} \sinh\left(\frac{2\omega d^{(2)} |\mathcal{A}_2|}{nc} z\right), \quad (10)$$

where $\mathcal{A}_1(0,\varphi)$ is constant and accounts for the initial conditions. Therefore the solution becomes:

$$A_1(\mathbf{r}, z) = \mathcal{A}_1(0, \varphi) e^{il\varphi} \cosh\left(\frac{2\omega d^{(2)} |\mathcal{A}_2|}{nc} z\right) + i\mathcal{A}_1^*(0, \varphi) e^{i(m-l)\varphi} \sinh\left(\frac{2\omega d^{(2)} |\mathcal{A}_2|}{nc} z\right). \quad (11)$$

The OAM of the fundamental field expressed by Eq. (11) can be estimated by making use of what was found in Eq. (4), considering the two terms of $A_1$ as the fields of the linear superposition. Substituting separately the two terms of Eq. (11) in equation (1), it can be clearly seen that the OAM per photon is *l* for the first term, while is *m-l* for the second and remains constant during the propagation.

As the field propagates inside the crystal, for small values of z the hyperbolic cosine is dominant and the overall momentum equals *l*, i.e. the starting OAM of the FF; when z increases, the second term of Eq. (11) grows until it reaches a value comparable with the first one. Therefore the momentum of the fundamental field, after an appropriate length of interaction becomes *m*/2, namely half of the OAM of the second harmonic field, in order to fulfil the momentum conservation. In figure 1is plotted the behaviour of the OAM per photon in $\hbar$ units of the FF field with the propagation distance z for different starting values of the angular momenta of the SH and FF beams. It is evident how, regardless of the starting value, the momentum of the FF changes in order to match half of the momentum of the SH. Looking at equation (11) emerges that the distance over which the OAM changes is strictly dependent on the second harmonic input amplitude and on the nonlinear coefficient. This means that higher amplitudes (or higher nonlinearities) imply faster variations of the OAM.

**FIGURE 1**

## 4. Numerical Simulations

*4.1 Model*

In order to verify what was predicted in the previous section by the analytical solution of the seeded second harmonic generation process, we performed numerical simulations, solving via a beam propagation method the coupled equations.

Starting from the nonlinear wave equation, we rearranged the coupled equations in a more suitable fashion writing them in Cartesian coordinates; in this way, numerical integration can be easily carried out:

$$\frac{\partial A_1(x,y,z)}{\partial z}$$
$$= \frac{i}{2n_b k_0}\frac{\partial^2 A_1(x,y,z)}{\partial x^2} + \frac{i}{2n_b k_0}\frac{\partial^2 A_1(x,y,z)}{\partial y^2} + \frac{i}{2n_b}k_0\left[\left(n_\omega^2(x,y,z)-n_b^2\right) + 4d^{(2)}(x,y,z)\frac{A_1^*(x,y,z)A_2(x,y,z)}{A_1(x,y,z)}\right]A_1(x,y,z) \quad (12)$$
$$\frac{\partial A_2(x,y,z)}{\partial z}$$
$$= \frac{i}{4n_b k_0}\frac{\partial^2 A_2(x,y,z)}{\partial x^2} + \frac{i}{4n_b k_0}\frac{\partial^2 A_2(x,y,z)}{\partial y^2} + i\frac{k_0}{n_b}\left[\left(n_{2\omega}^2(x,y,z)-n_b^2\right) + 2d^{(2)}(x,y,z)\frac{A_1^2(x,y,z)}{A_2(x,y,z)}\right]A_2(x,y,z)$$

Where ω is the FF frequency, 2ω is the SH frequency, $d^{(2)}$ is the nonlinear coefficient, $n_\omega$ and $n_{2\omega}$ are respectively the spatial dependent refractive indices at the frequencies ω and 2ω, $n_b$ is the background refractive index, $k_0$ is the vacuum wave vector and $A_j(x,y,z)$ is the complex envelope of each interaction field. The last term in Eqs. (12) can be simplified by using $A_j = |A_j(x,y,z)|\exp(-i\phi_j(x,y,z))$ being $j$ the index related to each field; in this way, in the first equation, the dependence from the input fields and from their relative phases becomes evident. Looking at the second of Eqs. (12) it can be noticed that, when the SH input signal is more intense than the FF input beam, the last term in the square brackets is small if compared to the linear part of the equation; thus the SH behaves essentially like in a non depletion case. On the other hand, in the first of Eqs. (12) the intense SH field enhances the nonlinear term, resulting in a strong nonlinear dependence of the FF. Moreover a strong relative input phase dependence of the two signals is also expected.

In our numerical analysis of the nonlinear interactions, we directly solve Eqs. (12) using a beam propagation method based on a fast Fourier transform (FFT) with a 200x200 mesh in the transverse plane.

*4.2 Input conditions*

In all the simulations we solved the equations considering a FF and a SH field, respectively with a wavelength of 800 nm and 400 nm, both propagating collinearly in a type I nonlinear crystal with a nonlinearity $d^{(2)}$ of 50 pm/V; the same effects considered in our analysis can also be obtained by selecting a crystal with different nonlinear coefficient, as BBO or LiNbO3 [24], adjusting the total crystal length and the SH intensity level. Furthermore we assume a beam waist of the FF of the order of 10 μm and of the SH reduced of a factor $\sqrt{2}$ with respect to the fundamental field.

In order to emphasize the role of the intensity of the SH field in controlling the OAM of the FF beam, several simulations were carried out, changing in each one the intensity level of the SH beam and choosing them sufficiently high, with respect to the FF, in order to avoid energy transfers from the fundamental to the second harmonic field.

Furthermore we investigated three cases, letting interact fields with different combinations of OAMs, as happened in figure 1. In the first situation, at the input plane of the nonlinear crystal there are a FF and SH fields with angular momentum per photon respectively equal to 0 and 2. In the second case the SH field carries zero angular momentum while the fundamental field has an OAM $\ell = 1$ and in the last one, both field carry an orbital angular momentum with the fractional value of 0.5.

In the first two cases, when the interacting fields possess integer OAMs, we used purely azimuthal Laguerre – Gaussian modes ($LG_{l0}$), i.e. with zero radial index; the OAM per photon of such beams is in fact equal to the azimuthal index $l$ [5]. On the other hand, when we deal with fractional OAMs (third situation), we modelled our fields by displacing an integer phase singularity with unitary topological charge from the beam's axis [25] in order to obtain an angular momentum per photon of 0.5 in $\hbar$ units. As a matter of fact, reference [25] shows that the intrinsic OAM of a beam with a displaced OV reduces through a Gaussian law with the extents of the displacement: $\ell \propto Q \exp(-2 r_v^2/w_0^2)$, being $r_v$ the displacement of the singularity, $w_0$ the spot size and Q the topological charge.

In this paper we limited our analysis of the fractional case considering the interaction of a FF field and its corresponding collinear SH generated beam; according to the Gaussian dependence of the OAM predicted in reference [25], it can be easily found for a fundamental field with $\ell = 0.5$ that, because of the momentum conservation, its collinear SH field should possess an OAM of $0.5\hbar$ per photon, that is the same as FF. Therefore this situation is still consistent with the hypotheses of equal fractional OAMs which brought us from Eq. (3) to Eq. (4).

The transverse input beam profiles $u(\rho,\varphi)$ of the FF fields are reported in fig. 2 (first row), together with the corresponding SH input profiles (second row) considered in the simulated cases. Figure 3 shows the phase profiles of the interacting fields of figure 2: when OAM is nonzero the phase winds around a singularity; in figure 3b can be noticed the displaced OV and the doubled topological charge of its corresponding SH field.

**FIGURE 2**

**FIGURE 3**

The off axis OV beam were modelled via a LG decomposition, using eq. (2); these fields were numerically evaluated by integrating the coefficients of the superposition with a 200x200 mesh, over a distance 5 times larger than the beam waist.

In all of the three situations the numerically evaluated transverse fields' profiles were propagated in the nonlinear crystal using the BPM in order to solve directly Eqs. (12). Equation (11) predicts that the length over which the change in OAM happens is strongly dependent on the amplitude of the SH field. Therefore, in order to keep the computational requirements low, i.e. reasonably short propagation distances, we performed our simulations by taking both the FF and SH input beams with very high amplitudes, of the order of 10^6 V/m to 10^7 V/m. Clearly it is possible to consider lower intensities (or higher nonlinearities) provided that larger propagation distances, i.e. longer crystals, are chosen.

*4.3 Results*

Looking at Eqs. (12), in phase matching conditions and with a relative input phase of $-\pi/2$, an amplification process of the FF occurs. In our simulations we followed the propagation of the fields over a length of 300 μm with a SH amplitude ranging from about 10 to 100 times higher than the FF, keeping the propagation away from the spatial soliton threshold.

In Fig. 4 the variation of the value of the fundamental field's OAM per photon in $\hbar$ units with the propagation distance are plotted for different levels of the SH amplitude. In all the considered cases can be seen that the OAM of the FF field varies with an hyperbolic dependence on the propagation distance, until it reaches half of the OAM of the second harmonic beam; on the other hand the momentum of the SH remains unaffected. As a matter of fact the SH input signal is so intense to propagate almost undepleted, while the FF beam undergoes amplification.

It can be noticed that these results are in perfect agreement with what was predicted in the previous section (fig. 1). Again, from figure 4, emerges that the distance over which a change in the value of OAM of the FF manifests depends on the amplitude level of the SH. Of course higher amplitudes of the SH input field result in larger amplification of the fundamental beam with a consequent faster variation of the momentum.

**FIGURE 4**

The change in the OAM due to the nonlinear interaction affects also the spatial transverse profile of the FF field. Figures 5 and 6 show respectively the intensity and phase profiles of the FF fields in the three cases after different distances of propagation. In particular, looking at the second situation where the final OAM tends to zero, it is possible to notice in the field's distribution the absence of the rotation, associated with the propagation, which is typical of fields with zero OAM [7]. On the contrary, in the other cases where the OAM is different from zero, the field's profile rotates as the propagation distance increases. In these situations the field's distribution is found, after 300 μm of propagation, rotated of a larger amount with respect to the second case. Looking at figure 6a it can be noticed that, as the OAM changes from zero to its ending value, OVs are created and, as the field propagates, the phase begins to whirl around the singularities, as also happens in fig 6c; on the other hand (fig 6b) a zero ending OAM implies the splitting of the initial vortex distribution and leads to a nonrotating phase profile.

**FIGURE 5**

**FIGURE 6**

Therefore the effect of the SH field is to modify the propagation length along which the OAM value changes from its starting to its ending value. In this way, with a fixed crystal length, by properly choosing the SH intensity level, it is possible to tune the output OAM of the FF field at one of the possible intermediate values between its starting value and half of the SH angular momentum.

If a non-phase matched process is considered (in this case we selected a coherence length of 1 μm, taking $n_\omega$=2.7 and $n_{2\omega}$=2.9) we obtain a behaviour of the FF field distributions rather different from the phase matched case. In Fig. 7 is showed the evolution of the OAM with the propagation length in these conditions: the behaviour of the fields in the nonlinear interaction is independent from the initial phase difference among the FF and SH beams and, as is showed in Fig. 7 for the case of FF with unitary OAM, the orbital angular momentum of the fundamental field is held constant during the propagation distance .

**FIGURE 7**

## 4. Conclusions

In the previous sections we analyzed the evolution of the OAM in a SH process by analytically solving the coupled nonlinear equations under the undepleted approximation for the second harmonic field. The results show that, as the interaction takes place, the orbital angular momentum of the field at frequency ω changes, tending to reach half of the SH momentum over a distance controlled by the amplitude of the SH beam. Moreover we studied numerically the same nonlinear interaction and found an OAM's behaviour adherent to what predicted theoretically. Furthermore we showed that the amplitude of the SH field in phase matching conditions is a parameter which grants control over the OAM of the fundamental field, allowing, with a fixed crystal's length, to obtain a field with a certain orbital angular momentum per photon, always conserving the total angular momentum. We then solved numerically Eqs. (12) when there is no phase matching and showed that, in this situation, is not possible to control the OAM. Therefore, using a seeded second harmonic phase-matched interaction, it is possible to reach a fast and efficient control of the orbital angular momentum of a field at frequency ω, by simply changing the momentum of the input SH field and operating on a parameter such as the input field's amplitude.

**Figures Captions**

Fig. 1. OAM per photon of the fundamental field vs. propagation length inside the nonlinear crystal for different input values of the SH and FF OAMs and different amplitudes of the SH field; Blue: $7\times10^6$ V/m, Red: $2\times10^7$ V/m, Black: $4\times10^7$ V/m;. a) FF with $\ell = 0$, SH with $\ell = 2$; b) FF with $\ell = 1$, SH with $\ell = 0$; c) FF with $\ell = 1/2$, SH with $\ell = 1/2$. During the interaction the FF field changes its OAM from its starting value to half of the OAM of the SH field.

Fig. 2. Intensity profiles of the interacting fields at the input plane. First row: FF input beam; second row: SH input beam. a) FF with $\ell = 0$, SH with $\ell = 2$; b) FF with $\ell = 1$, SH with $\ell = 0$; c) FF with $\ell = 1/2$, SH with $\ell = 1/2$.

Fig. 3. Phase profiles of the interacting fields at the input plane. First row: FF input beam; second row: SH input beam. a) FF with $\ell = 0$, SH with $\ell = 2$; b) FF with $\ell = 1$, SH with $\ell = 0$; c) FF with $\ell = 1/2$, SH with $\ell = 1/2$.

Fig. 4. OAM per photon (in $\hbar$ units) vs. propagation distance in the nonlinear crystal for a FF with an amplitude of $3\times10^5$ V/m and different input amplitudes of the second harmonic field. Blue: $7\times10^6$ V/m, Red: $2\times10^7$ V/m, Black: $4\times10^7$ V/m; a) FF with $\ell = 0$, SH with $\ell = 2$; b) FF with $\ell = 1$, SH with $\ell = 0$; c) FF with $\ell = 1/2$, SH with $\ell = 1/2$.

Fig. 5. Intensity profiles of the three FF fields of figure 2 after 30 μm (first row) and 300 μm (second row) of propagation with high SH input amplitude: after an appropriate propagation length the OAM per photon of the FF is half of the SH's momentum.

Fig. 6. Phase profiles of the three FF fields after 30 μm (first row) and 300 μm (second row) of propagation with high SH input amplitude: the OAM is bound to the presence of OVs. In the second case the final OAM is 0 and an edge dislocation is present.

Fig.7. OAM per photon (in $\hbar$ units) of the FF and SH fields when $\ell_{FF} = 1$ and $\ell_{SH} = 0$: the OAM of the FF remains constant during the propagation.

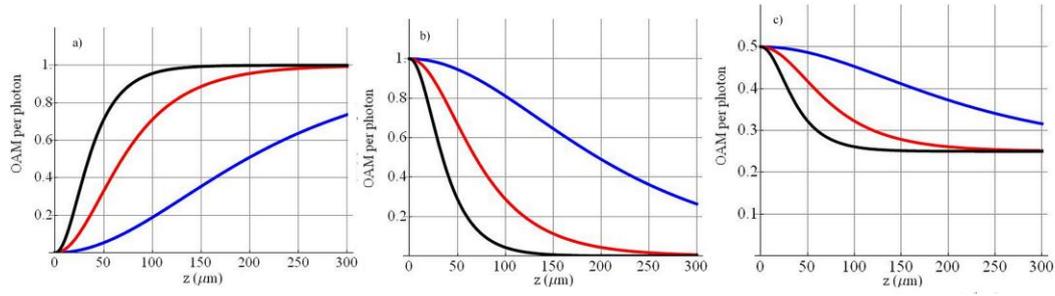

FIGURE 1

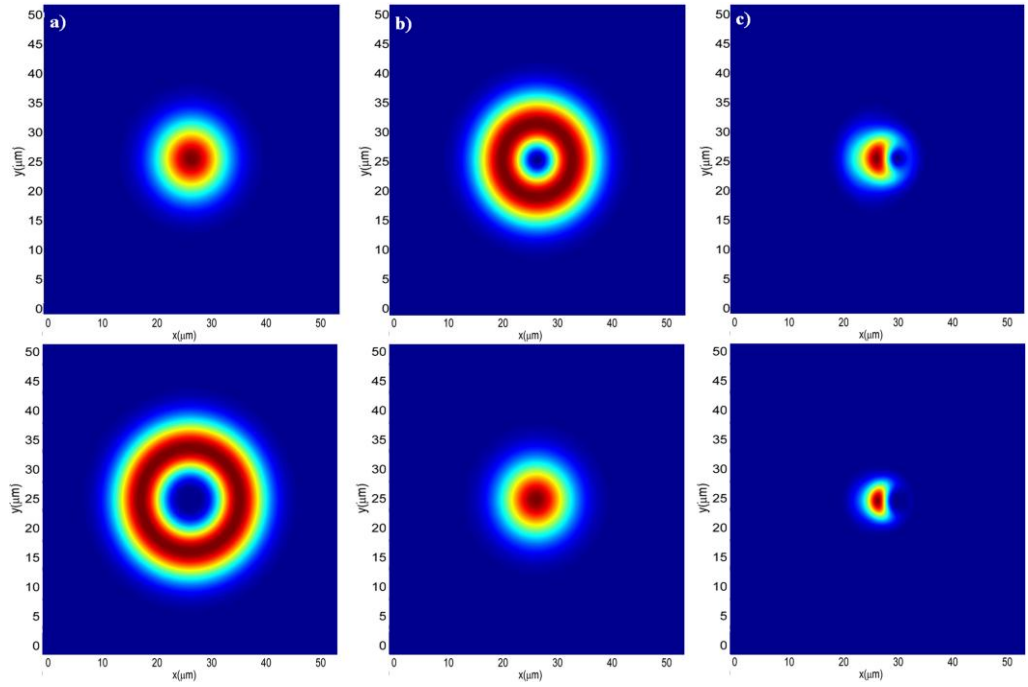

FIGURE 2

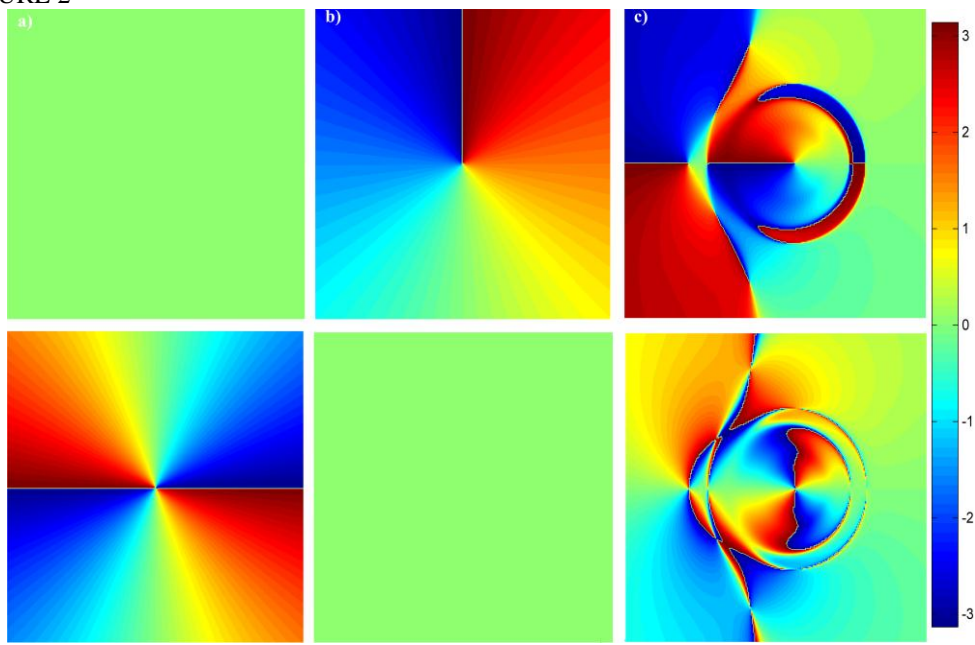

FIGURE 3

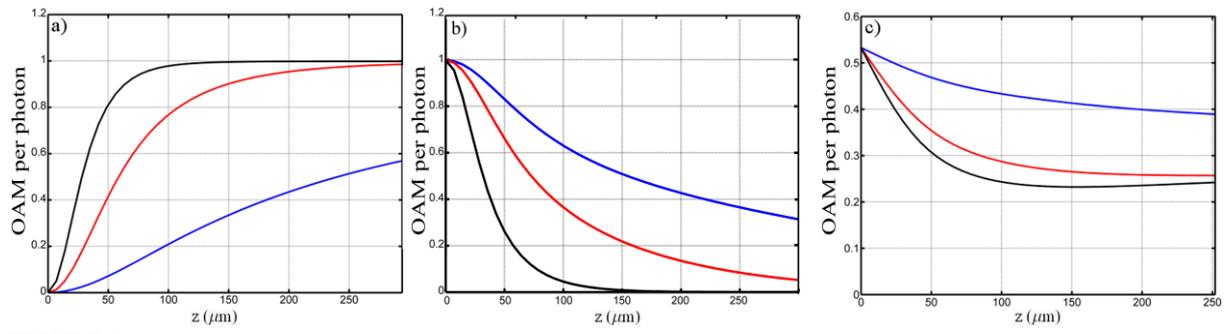

FIGURE 4

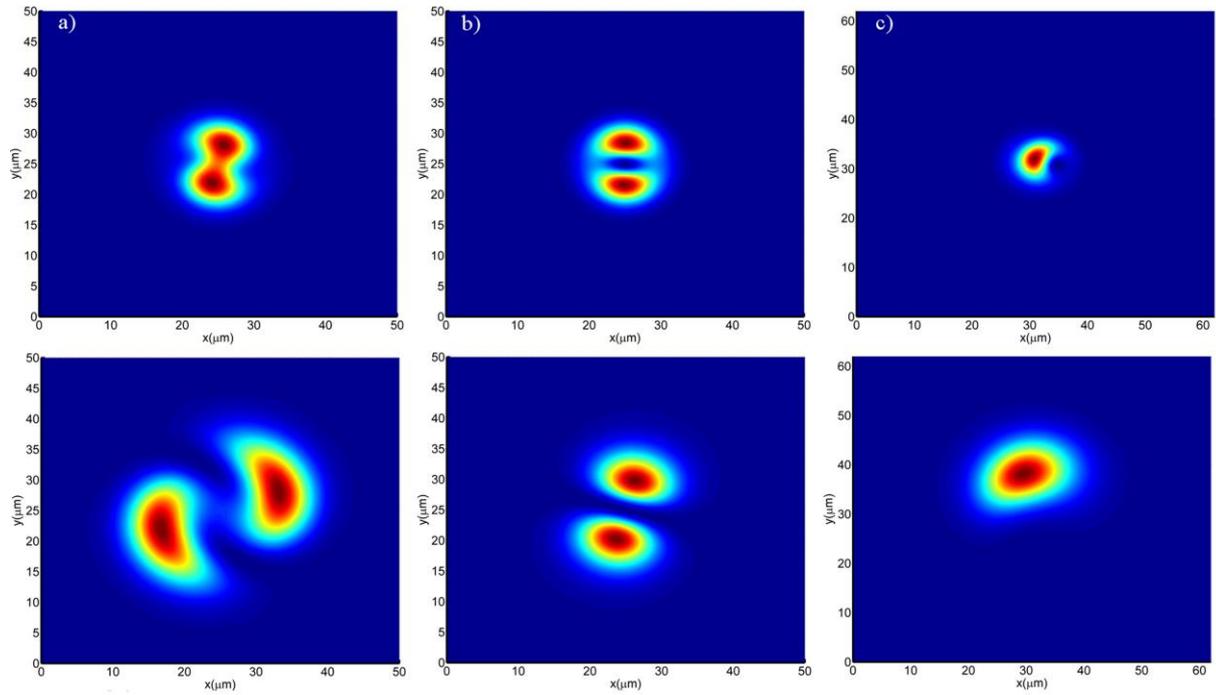

FIGURE 5

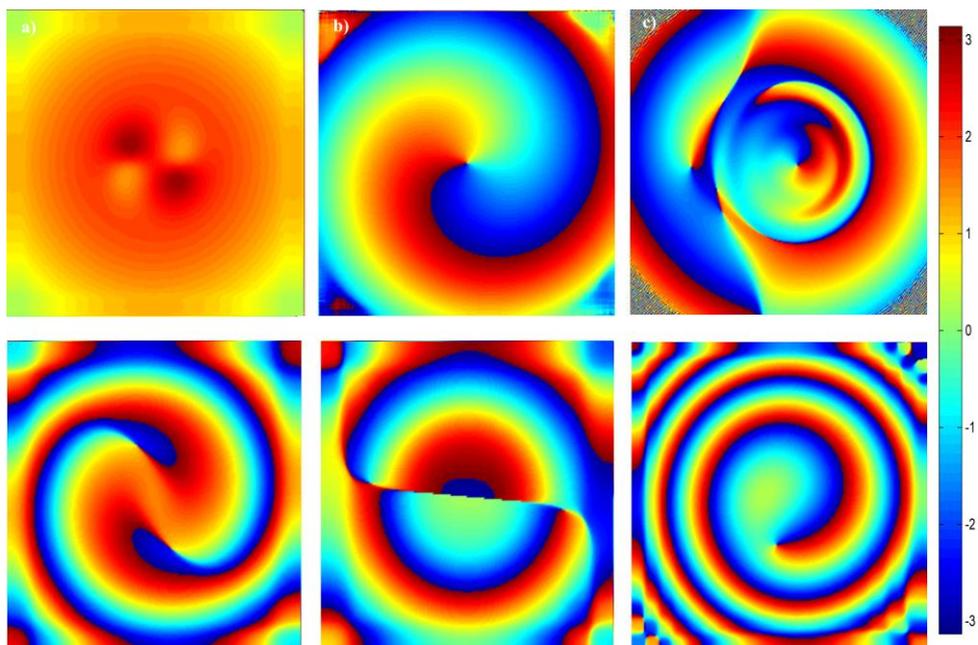

FIGURE 6

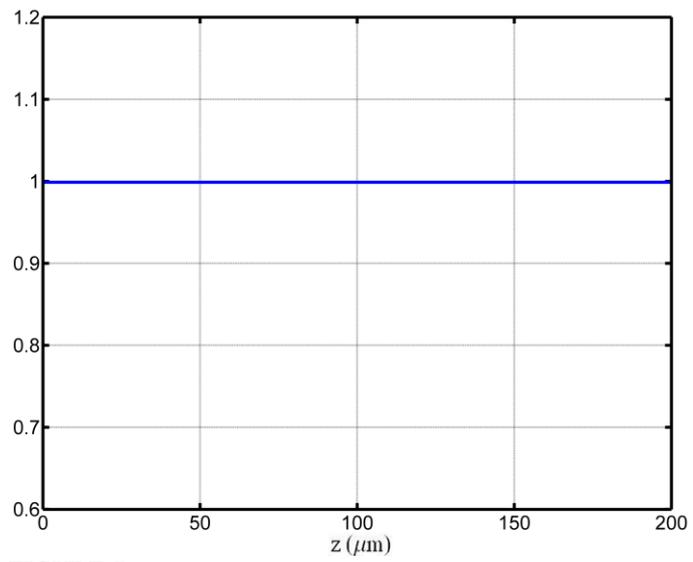
FIGURE 7